\documentclass[prl,floatfix,twocolumn,showpacs]{revtex4}

\usepackage{graphicx}

\renewcommand{\vec}[1]{\mbox{\boldmath$#1$}}
\def\q{\qquad}
\def\beg{\begin{eqnarray}}
\def\ende{\end{eqnarray}}
\def\gsim{\lower.4ex\hbox{$\;\buildrel >\over{\scriptstyle\sim}\;$}} 
\def\lsim{\lower.4ex\hbox{$\;\buildrel <\over{\scriptstyle\sim}\;$}} 

\def \Re {\mathrm{Re}}
\def \Rm {\mathrm{Rm}}
\def \Ha {\mathrm{Ha}}
\def \Pm {\mathrm{Pm}}
\def \muo{{\mu}_\Omega}

\begin{document}
\title{Turbulent magnetic Prandtl numbers obtained with  MHD Taylor-Couette flow experiments}
\author{Marcus Gellert \&  G\"unther R\"udiger}
\affiliation{Astrophysikalisches Institut Potsdam,
         An der Sternwarte 16, D-14482 Potsdam, Germany}
\email{ mgellert@aip.de, gruediger@aip.de }

\date{\today}
\begin{abstract}
The  stability problem of   MHD Taylor-Couette flows
with toroidal  magnetic fields  is considered  in dependence on the  magnetic Prandtl number. 
Only the most  uniform (but not current-free) field  with  $B_{\rm in}=B_{\rm out}$  has been considered.  For high  enough Hartmann numbers the toroidal field 
is always unstable. Rigid rotation, however, stabilizes the magnetic (kink-)instability.

The axial current which drives the instability is reduced by the  electromotive force induced by the instability itself. Numerical simulations are presented 
to probe this effect as a possibility  to measure the turbulent conductivity  in a laboratory. It is shown numerically that  in a  sodium  experiment 
(without rotation) an eddy diffusivity 4 times the molecular diffusivity  appears resulting in a potential difference of $\sim 34$ mV/m.
If the cylinders are rotating then also the eddy viscosity can be measured. Nonlinear simulations of the instability lead to a turbulent magnetic 
Prandtl number of 2.1 for a molecular magnetic Prandtl number of $0.01$. The trend goes to higher values for smaller Pm. 
\end{abstract}

\maketitle
%%%%%%%%%%%%%%%%%%%%%%%%%%%%%%%%%%%%%%%%%%%%%%%%%%%%%%%%%%%%%%%%%%%%%%
\section{Introduction}
%%%%%%%%%%%%%%%%%%%%%%%%%%%%%%%%%%%%%%%%%%%%%%%%%%%%%%%%%%%%%%%%%%%%%%%

Strong enough toroidal fields that are not current-free  become unstable  due to  the Tayler instability (TI,  \cite{T57,V72,T73}).  Because the source of the
energy is  the electric current, these (mainly nonaxisymmetric) instabilities can exist even without any 
rotation.  On the other hand, it is becoming increasingly clear that the stability of differential rotation under the presence of  magnetic fields is 
one of the key problems in MHD astrophysics. There, however,  is no laboratory experiment so far and even the related numerical simulations of TI are 
very  rare \cite{BRW06,GRE08}.  We shall demonstrate here  how the TI  interacts with differential rotation, 
and how it is possible to  verify  the main results in
laboratory experiments.  In particular, the theoretical results are used  to propose    experiments for measuring the turbulent diffusivity   via a  
TI-induced reduction of the electromotive force. Such  experiments will be of high relevance as the knowledge of the magnetic turbulent  diffusivity is 
basic for many applications in fluid dynamics. Very often  we have only  limited informations about the magnetic diffusivity. 
In the laboratory only a very small number of  experiments have been done (see \cite{RB01,Frick07}). The same is true for the eddy 
viscosity which can be measured with the same experimental device so that finally the turbulent magnetic Prandtl number becomes known for one and 
the same instability.

Consider a Taylor-Couette (TC) flow with   $\vec{U}$ as the velocity, $\vec{B}$ the magnetic field,  $\nu$ the microscopic kinematic viscosity and 
$\eta$ the microscopic magnetic diffusivity.
The  basic state in cylindrical geometry is $U_R=U_z=B_R=B_z=0$ and
\beg
U_\phi=R\Omega=a_\Omega R+\frac{b_\Omega}{R},  \q
B_\phi=a_B R+\frac{b_B}{R}.
\label{basic}
\ende
Let 
\beg
\hat\eta=\frac{R_{\rm{in}}}{R_{\rm{out}}}, \; \; \;
\mu_\Omega=\frac{\Omega_{\rm{out}}}{\Omega_{\rm{in}}},  \; \; \;
\mu_B=\frac{B_{\rm{out}}}{B_{\rm{in}}}.
\ende
$R_{\rm{in}}$ and $R_{\rm{out}}$ are the radii of the inner and outer cylinders, $\Omega_{\rm{in}}$ and $\Omega_{\rm{out}}$  their rotation
rates  and $B_{\rm{in}}$ and $B_{\rm{out}}$ are the azimuthal magnetic fields at the inner and outer cylinders.
In particular, a field of the form $b_B/R$ is generated by  an axial current only through the inner region $R<R_{\rm{in}}$, whereas a
field of the form $a_B R$ is generated by  a uniform axial current through the entire region $R<R_{\rm{out}}$ including the fluid.  

The magnetic Prandtl number $\Pm$, the Reynolds number $\Re$ and the Hartmann number $\Ha$, 
\beg
{\rm Pm}=\frac{\nu}{\eta}, \quad
{\rm{Re}}=\frac{\Omega_{\rm{in}} R_0^2}{\nu}, \quad
{\rm{Ha}}=\frac{B_{\rm{in}} R_0}{\sqrt{\mu_0 \rho \nu \eta}},
\ende
are the basic parameters of the problem where $R_0=\sqrt{D R_{\rm in}}$ is the unit of length with $D=R_{\rm{out}}-R_{\rm{in}}$.
 For the velocity the boundary conditions are assumed as always no-slip ($\vec{u}=0$). For conducting walls the radial component
of the field and the tangential components of the current must vanish so that 
$
{\rm d}b_\phi/{\rm d}R + b_\phi/R = b_R = 0
$
 at both $R_{\rm{in}}$ and $R_{\rm{out}}$. Here  $\vec{u}$ and $\vec{b}$ are the fluctuating components of flow and field.

While the linear stability code works well with small $\Pm$, the minimum microscopic $\Pm$ which can be handled with our nonlinear code is 
(only) $10^{-2}$. The numerics  cannot deal with the very small magnetic Prandtl numbers of liquid metals used in the laboratory ($\rm Pm \lsim 10^{-5}$). 
Some of our results can only be obtained  by extrapolation methods. 

%%%%%%%%%%%%%%%%%%%%%%%%%%%%%%%%%%%%%%%%%%%%%%%%%%%%%%%%%%%%%%%%%%%%%%%%%%%
\section{The instability map}
%%%%%%%%%%%%%%%%%%%%%%%%%%%%%%%%%%%%%%%%%%%%%%%%%%%%%%%%%%%%%%%%%%%%%%%%%%%%
The  map for the marginal   instability is the result of a linear theory for liquid sodium with $\Pm=10^{-5}$. The linearized equations  of the MHD system in a TC flow under the presence of a toroidal 
field are  given  elsewhere, \cite{RHSE07}.

\begin{figure*}[htb]
\hbox{
\includegraphics[width=5.8cm,height=4.4cm]{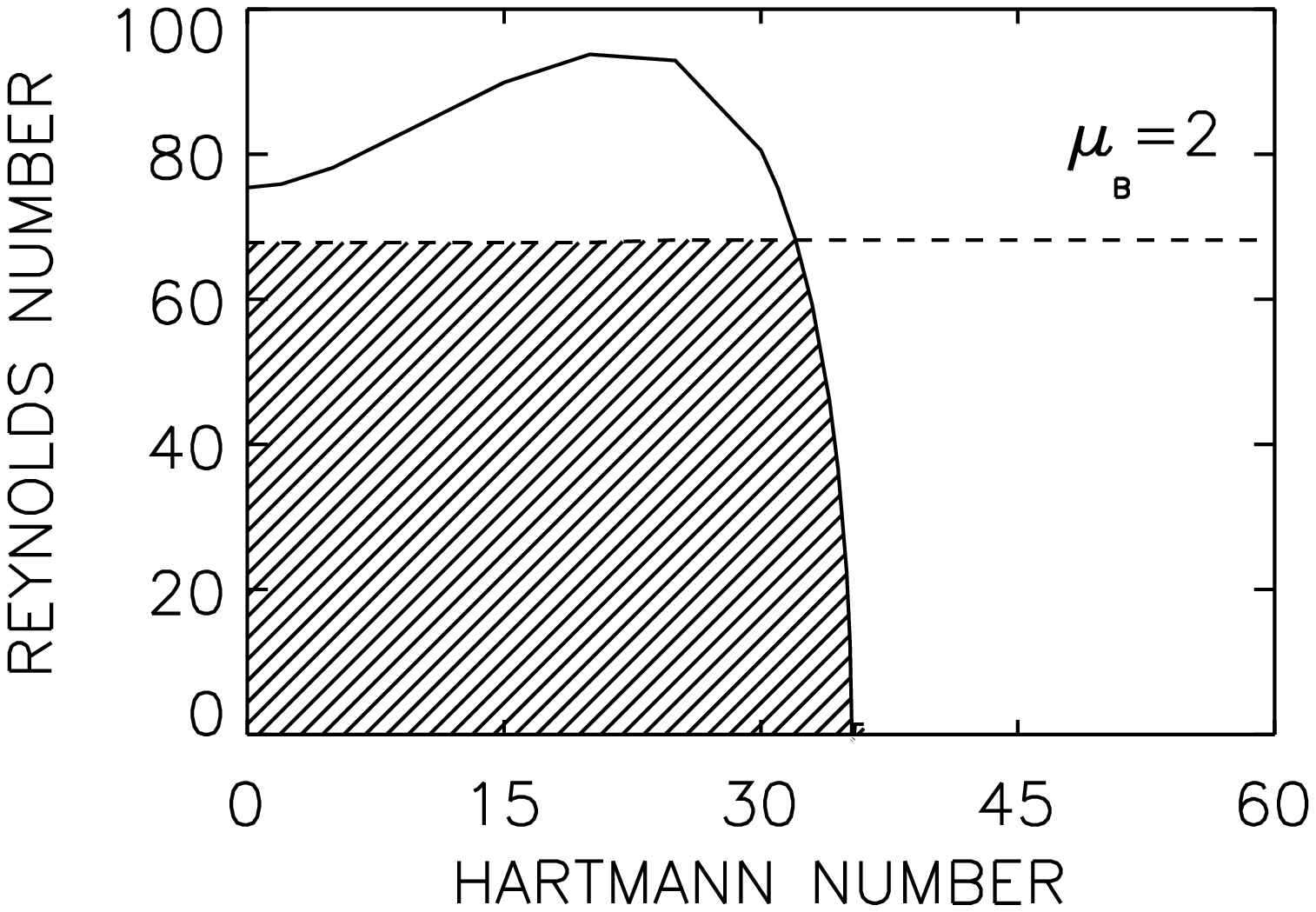}
\includegraphics[width=5.8cm,height=4.4cm]{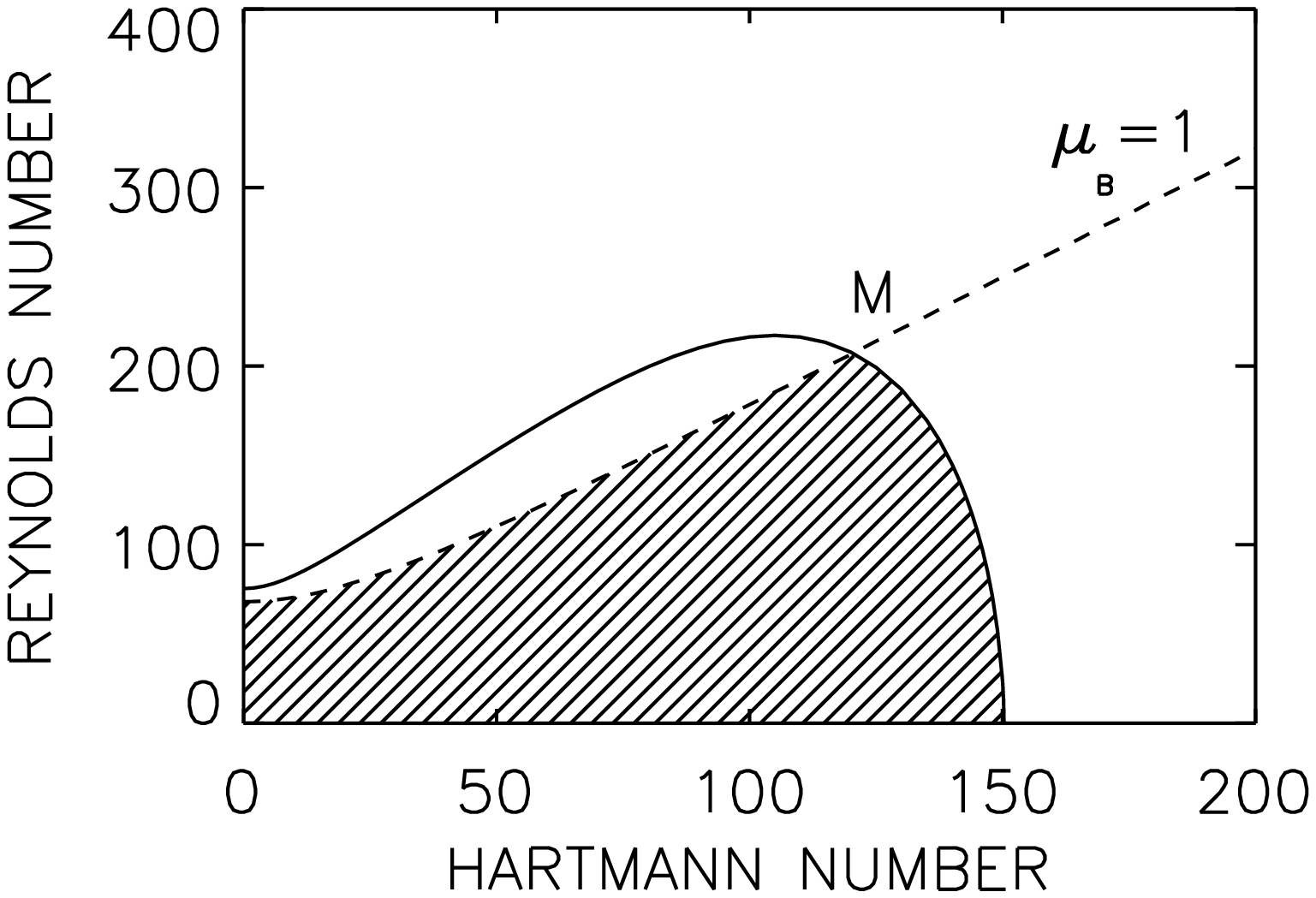}
\includegraphics[width=5.8cm,height=4.4cm]{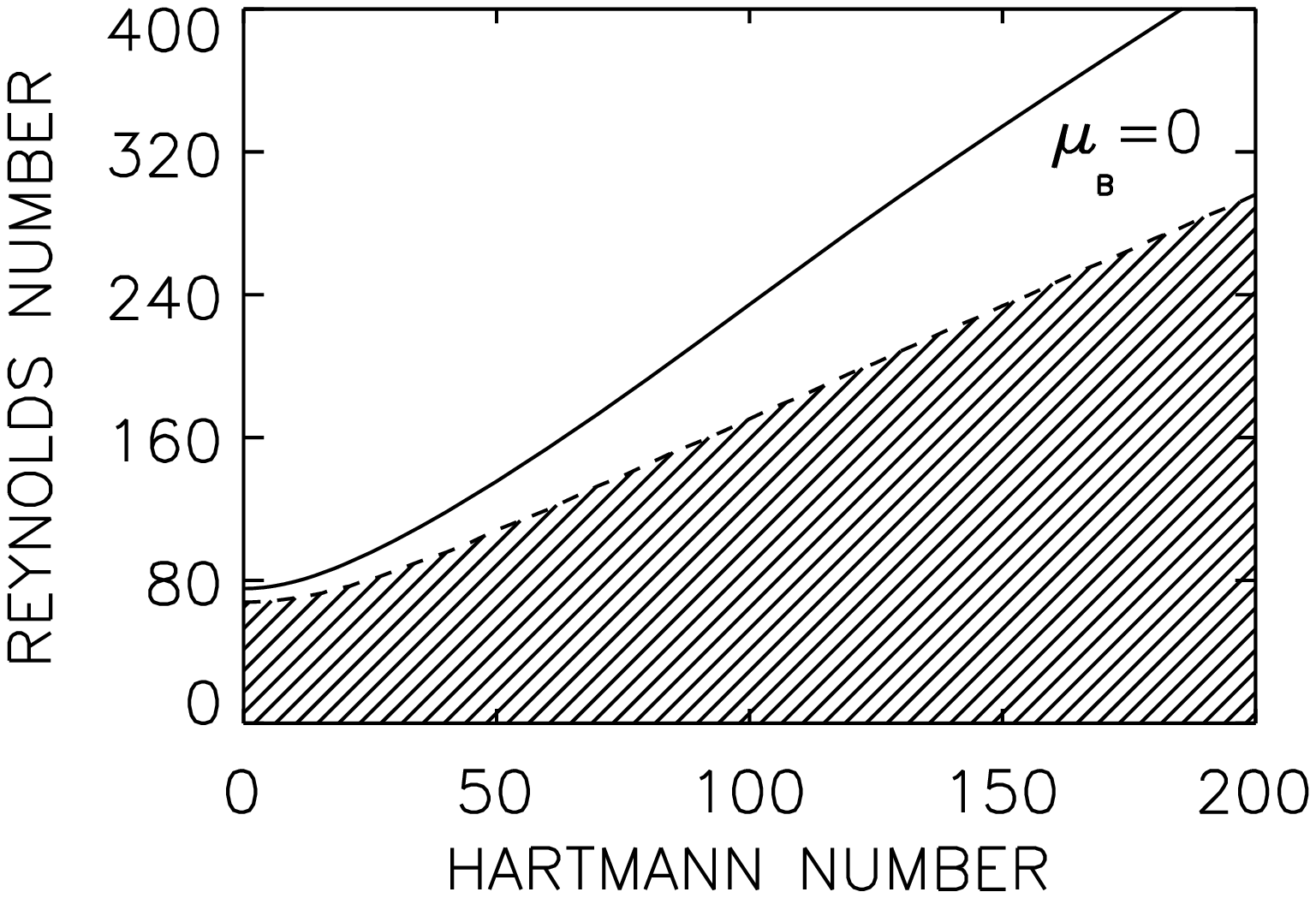}
}
\caption{\label{fig1} The marginal stability curves for $m=0$ (dashed)
         and $m=1$ (solid).  Pm$=10^{-5}$, $\hat\eta=0.5$, $\mu_\Omega=0$,
         and $\mu_B$ as indicated. Conducting walls.  The hatched  domain is stable.}
\end{figure*}

 Figure \ref{fig1} shows the results for various values of $\mu_B$.  For $\mu_B=0.5$ the toroidal field is current-free between the cylinders.
The outer cylinder is assumed as resting so that for vanishing magnetic field the rotation law is centrifugally-unstable for $\rm Re>68$. 
From the given profiles  closest to being current-free is $\mu_B=0$ and we do not find that  for
$\Ha\leq 200$ there is any  sign of destabilizing influence of the magnetic field, for neither axisymmetric nor nonaxisymmetric perturbations.

For certain $\mu_B$ the $m=1$ mode should be unstable while the $m=0$ mode should be stable \cite{S06}.
The values $\mu_B=1$ and $\mu_B=2$ in Fig. \ref{fig1} are examples of this situation. There is always a crossover point at which the most unstable mode changes
from $m=0$ to $m=1$.  Note also that for $\mu_B=1$ the critical Reynolds number  for the $m=0$ mode steadily increases, while the 
$m=1$ mode is  suddenly decreasing for a sufficiently strong magnetic field.  Hence, weak fields initially can stabilize the flow, and  stronger 
fields eventually destabilize  via a nonaxisymmetric mode.  Beyond $\Ha=150$  the flow is unstable even for $\Re=0$.

Except for the almost current-free profile $\mu_B=0$ all other values share the feature that there is a critical Hartmann number beyond which
the basic state is unstable even for $\Re=0$. Let Ha$^{(0)}$ and Ha$^{(1)}$ denote these critical Hartmann numbers for $m=0$ and $m=1$, resp.
  
%%%%%%%%%%%%%%%%%%%%%%%%%%%%%%%%%%%%%%%%%%%%%%%%%%%%%%%%%%%%%%%%%%%%%%%%%%%%%%%%%%%%%%%%%%%%%%%%%%%%%%%%%
\subsection{Flat rotation laws}
\begin{figure}[htb]
\includegraphics[width=8.5cm,height=5.5cm]{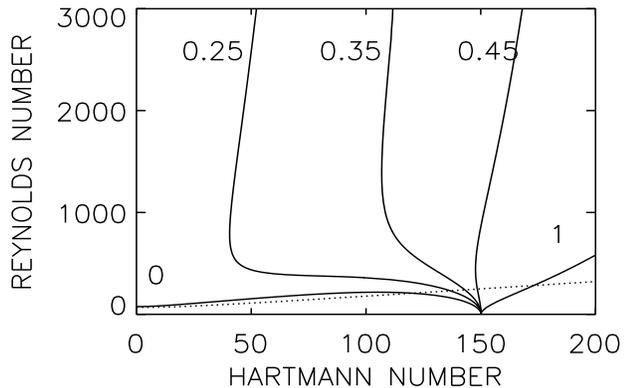}
\caption{\label{fig2} $\mu_B=1$. The same magnetic constellation as in Fig. \ref{fig1} (middle) but for  $\muo$ as indicated. $\muo=0$ (resting 
         outer cylinder), $\muo=0.25$ (Rayleigh limit), $\muo=0.35$, $\muo=0.45$, $\muo=1$ (rigid rotation).  
         The dotted line gives the eigenvalues for $m=0$  which for those  parameters only exists for  $\muo=0$.}
\end{figure}

The rotation law with resting outer cylinder destabilizes the magnetic field. The question arises what happens for those  flat rotation laws 
which are stable in the nonmagnetic regime. In Fig. \ref{fig2} the marginal stability curves are also given for $\muo=0.25$ 
(Rayleigh limit), $\muo=0.35$, $\muo=0.45$  and $\muo=1$ (rigid rotation). One finds the instabilities more and 
more {\em stabilized} by the rotation. 

Note the massive quenching of the TI by rigid rotation. Even a rather slow rotation prevents the TI to destabilize the system. Rigidly 
rotating containers   can keep   much stronger  fields as stable than without rotation. This rotational 
stabilization is  modified  for nonuniform rotation. At the Rayleigh limit, where $\Omega \propto R^{-2}$, even a  slow  
rotation destabilizes the system while it is stabilized  for  fast rotation. Generally, fast rotation  stabilizes, slow rotation 
destabilizes.  At a Hartmann number  of (say) 50 and at the Rayleigh line one finds  for increasing  rotation rate the 
regimes: stable, unstable,  stable. The critical Reynolds numbers of the sequence  are $\sim 300$ and $\sim 1800$ which can easily be  
realized in the laboratory. A  similar  situation holds for the (quasi-Kepler) rotation   law with  $\muo=0.35$ while 
for rotation laws with $\muo\gsim 0.45$ only the rotational  stabilization can be  observed.

%%%%%%%%%%%%%%%%%%%%%%%%%%%%%%%%%%%%%%%%%%%%%%%%%%%%%%%%%%%%%%%%%%%%%%%%%%%%%%%
\subsection{Electric currents}
%%%%%%%%%%%%%%%%%%%%%%%%%%%%%%%%%%%%%%%%%%%%%%%%%%%%%%%%%%%%%%%%%%%%%%%%%%%%%%
For  experiments  the electric currents must not be too strong. As an upper limit currents with 10--15 kA  shall be considered. In order 
to translate the obtained critical Hartman numbers into amplitudes of electrical currents we apply our results to liquid sodium with  
a density of 0.92 g/cm$^3$, a microscopic magnetic diffusivity of 810 cm$^2$/s and a magnetic Prandtl number of $10^{-5}$. For 
gallium-indium-tin the necessary  currents are stronger by a factor of 3.15 (see \cite{RSSH07}).

Let $I_{\rm{axis}}$ be the axial current inside the inner cylinder and $I_{\rm{fluid}}$  the axial current through  the fluid (i.e. between inner and outer
cylinder). Then the toroidal field amplitudes at the inner and outer cylinders are
\beg
B_{\rm{in}}=\frac{I_{\rm{axis}}}{5R_{\rm{in}}}, \q
B_{\rm{out}}=\frac{(I_{\rm{axis}}+I_{\rm{fluid}})}{5R_{\rm{out}}},
\label{bi}
\ende
measured in cm, Gauss and Ampere. Expressing $I_{\rm{axis}}$ and $I_{\rm{fluid}}$ in terms of our dimensionless parameters one finds 
\beg
I_{\rm{axis}}=
5 {\rm Ha}\ \sqrt{\frac{\hat\eta\mu_0\rho\nu\eta}{1-\hat\eta}} , \quad \quad I_{\rm{fluid}}=\frac{\mu_B-\hat\eta}{\hat\eta} I_{\rm{axis}}.
\label{iaxis}
\ende
   Table \ref{t1} gives the electric currents needed to reach the {\em less} of
Ha$^{(0)}$ and Ha$^{(1)}$ for  $\hat\eta= 0.5$, and  $\mu_B$ ranging from $-2$ to 2 in each case.  Note that for
large $|\mu_B|$ the current  $I_{\rm{fluid}}$ approaches a  constant value. 

\begin{table}[h]
%\begin{ruledtabular}
\caption{\label{t1} Characteristic Hartmann numbers and electric currents for
a  sodium-container ($\hat\eta=0.5$) with conducting walls. The experiment with the almost uniform
field $\mu_B=1$  is indicated in bold.}
%\medskip
\begin{tabular}{lccccc}
\hline
$\mu_B$ & $\Ha^{(0)}$ & $\Ha^{(1)}$ & $I_{\rm{axis}}$ [kA] &  $I_{\rm{fluid}}$ [kA]\\
\hline
-2 & 19.8 & 24.8 & 0.807 & -4.04 \\
-1 & 59.3 & 63.7 & 2.42 & -7.25 \\
\hline
{\bf 1} & $\infty$ & {\bf 151} & {\bf 6.16} & {\bf 6.16}  \\
 2 & $\infty$ & 35.3 & 1.44  & 4.32   \\
\hline
\end{tabular}
\end{table}

The most interesting experiment is that with the almost uniform field $\mu_B=1$.
For a container with a  gap of $\hat\eta=0.5$, parallel currents of 6.16 kA  
are necessary along the axis and through the fluid.  The experiment  does not possess the weakest electric currents but both the currents are parallel and have the 
same amplitudes. Figure \ref{fig1} (middle) shows that in this case a crossing point M exists where the axisymmetric 
mode has the same characteristic Reynolds number and Hartmann number as the nonaxisymmetric mode with $m=1$. 
%%%%%%%%%%%%%%%%%%%%%%%%%%%%%%%%%%%%%%%%%%%%%%%%%%%%%%%%%%%%%%%%%%%%%%%%
\section{The eddy diffusivity}
%%%%%%%%%%%%%%%%%%%%%%%%%%%%%%%%%%%%%%%%%%%%%%%%%%%%%%%%%%%%%%%%%%%%%%%%
We now turn to  the mean-field concept  turbulent fluids of electrically conducting material.  It is 
known that the existence of turbulence in the fluid reduces the electric conductivity or -- with other words -- 
the fluctuations enhance the  magnetic diffusivity  called the turbulent magnetic  diffusivity. In MHD  the turbulent  diffusivity is a much more simple 
quantity than the corresponding eddy viscosity. While the latter is also formed by the existing magnetic fluctuations this is not 
the case for the turbulent  diffusivity. In a simplified (`SOCA') approximation  for a turbulence field with a 
correlation time $\tau_{\rm corr}$ results
\beg
\nu_T\simeq (\frac{2}{15} \langle u^2\rangle+ \frac{1}{3}  \frac{\langle b^2\rangle }{\mu_0\rho})\ \tau_{\rm corr}
\label{diff1}
\ende
for the eddy viscosity but only
\beg
\eta_T\simeq \frac{\tau_{\rm corr}}{3} \langle u^2\rangle
\label{diff2}
\ende
for the eddy diffusivity, \cite{VK83} . The magnetic fluctuations do not contribute to the magnetic diffusivity. This basic difference 
between both the diffusion coefficients is not yet proven by an experiment. The results (\ref{diff1}) and (\ref{diff2}) suggest that in turbulent magnetic 
fluids  the effective magnetic Prandtl number  exceeds the value 0.4 which was  confirmed   by   numerical 
simulations for driven MHD turbulence with Pm of order unity, \cite{YOUS03}. The knowledge of the turbulent magnetic Prandtl number is 
of extraordinary meaning in fluid mechanics and geo/astrophysics. For  its calculation one has to measure both quantities 
simultaneously in one and the same experiment.

Simplifying, the nonaxisymmetric components of  flow and  field may  be used in the following as the `fluctuations'  while the  
axisymmetric components are considered as the mean quantities. Then the averaging procedure is simply the integration over  
the azimuth $\phi$. 
It is standard to express the turbulence-induced electromotive force (EMF) as
\beg
\vec{\cal{E}}= \langle\vec{u} \times \vec{b}  \rangle =  - \eta_{\rm T}{\rm curl} \vec{B}
\label{veccalE}
\ende
with  the (scalar) eddy diffusivity $\eta_{\rm T}$ which must be positive. In  cylindric geometry the mean 
current ${\rm curl}\vec{B}$  has only a $z$-component. Hence, 
$$
{\cal E}_z=
- \eta_{\rm T}{\rm curl}_z \vec{B}.
$$ 

Take from Table \ref{t1} that for $\mu_B=1$ the current through the fluid is positive then  for {\em negative} ${\cal E}_z$ 
the $\eta_{\rm T}$   results as positive. This is indeed the case.  We 
have shown that  TI indeed provides reasonable expressions for the turbulent  diffusivity in  rotating containers, \cite{RGS08}.
%%%%%%%%%%%%%%%%%%%%%%%%%%%%%%%%%%%%%%%%%%%%%%%%%%%%%%%%%%%%%%%%%%%%%%%%%%%%%%%%%%%%%%%%%%%%%%%%%%%%
\subsection{Nonlinear simulations}
%%%%%%%%%%%%%%%%%%%%%%%%%%%%%%%%%%%%%%%%%%%%%%%%%%%%%%%%%%%%%%%%%%%%%%%%%%%%%%%%%%%%%%%%}
The   absolute values for ${\cal E}_z$ can only be computed with nonlinear simulations. The minimum possible magnetic Prandtl number for the code 
yielding robust results is  $10^{-2}$. Here the results without and with rotation but only for $\Ha=200$ 
are reported. The used MHD Fourier spectral element 
code has been  described earlier in more detail, \cite{FOURN05}, \cite{GRF07}. 

Either $M=8$ or $M=16$ Fourier modes are used, two or three elements in radius and twelve or eighteen elements in axial direction, 
resp. The polynomial order is varied between $N=8$ and $N=16$.

\begin{figure}[htb]
\includegraphics[width=8.0cm,height=5.0cm]{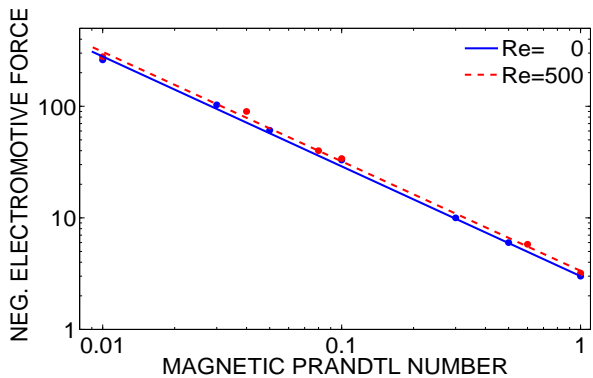}
%\caption{\label{fig5} The same magnetic constellation as in Fig. \ref{fig4} but for  the maximum values of the normalized axial EMF  
%         vs $\Pm$. The lower curve is without rotation, the  upper curve is with  rotation and shear ($\muo=0.35$).}
\caption{\label{fig5} Maximum values of the normalized axial EMF vs $\Pm$ for $\Ha=200$ and $\mu_B=1$. The lower curve is 
         without rotation, the  upper curve is with rotation and shear ($\Re=500,\muo=0.35$).}
\end{figure}

%We start with $\Ha=200$ and $\Re=0$. 
%Figure \ref{fig4} shows the TI-induced EMF  as negative with a maximum absolute 
%value of 260 (in units normalized with $\nu/D$ and $B_{\rm in}$). 
Figure \ref{fig5} shows the negative TI-induced EMF for magnetic Prandtl numbers varied between 0.01 to 1. The main results are    that i) the  EMF is always negative ($\eta_{\rm T}$ positive!) and ii) it runs with $E/ \Pm$ with the factor $E\simeq 3$ taken from the plot. 
The resulting EMF in physical units is  $-\eta E B_0/D$. Hence,  
 \beg
 \frac{\eta_{\rm T}}{\eta} \simeq 1.5 E \simeq 4.5.
 \label{etat}
 \ende
For  the voltage difference  $\delta U$ due to this EMF  one  finds
$
\delta U = \eta E B_0 H/D
$
with $H$ as  the container height and 
\begin{equation}
 \delta U=\frac{ \eta E \Gamma \sqrt{\mu_0\rho\nu\eta} {\rm Ha}}{D}
\end{equation}
with $\Gamma=H/D$ the aspect ratio of the container. For $D =10$ cm and $H=100$ cm we find  for 
sodium ($\sqrt{\mu_0 \rho \nu \eta} \simeq 8.15$) the maximum value of 34 mV as the potential difference from endplate 
to endplate. This  value can only be considered as an estimate 
basing on the scaling  with $1/\Pm$ suggested by Fig. \ref{fig5}. But even in the case that the 
slope of the curve decreases for smaller $\Pm$  the effect should be observable in the laboratory. A Hartmann 
number of 200 requires 163 G at the inner cylinder ($R_{\rm in}=10$ cm) which can be produced with an axial current of 8.15 kA for $R_{\rm out}= 2 R_{\rm in}$.

In order to study the rotational influence also the Pm-dependence of the EMF  under the presence of a differential rotation is 
given  in  Fig. \ref{fig5}. It is $\muo=0.35$ (quasikeplerian) and the Reynolds number is $\Re =500$. Note that the influence of the rotation 
for the  TI-induced EMF is surprisingly weak; with rotation the values  are slightly higher than without rotation.

\begin{figure}[htb]
\includegraphics[width=8.0cm,height=5.0cm]{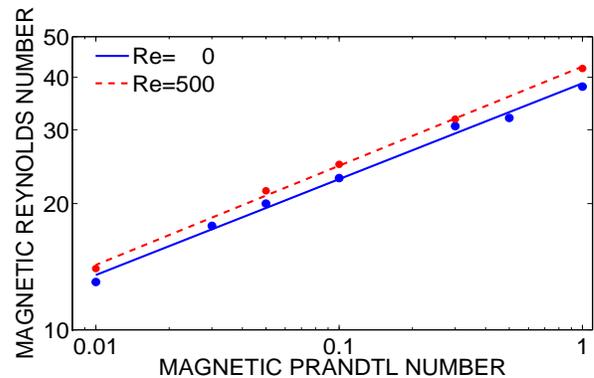}
\caption{\label{fig6} The same as in Fig. \ref{fig5} but for the magnetic Reynolds number of the fluctuations.}
\end{figure}

The  magnetic Reynolds number, $\Rm={\rm MAX}(u)D/\eta$, of the fluctuations is considered next. With  Fig. \ref{fig6} a  rather weak magnetic Prandtl number dependence of $\Rm$ is found. Extrapolating the results to $\rm Pm=10^{-5}$ 
 gives in both cases a value of $\Rm\simeq 2.6$. The associated velocity fluctuations for sodium are about  15 m/s in a gap of 1 cm and 1.5  m/s  in a gap of 10  cm. The values are rather similar to those of the Riga '$\alpha$-yashchik'  experiment, \cite{KR80}. 
Even with resting cylinders it is possible to produce rather high (azimuthal) velocities in TI experiments.

%%%%%%%%%%%%%%%%%%%%%%%%%%%%%%%%%%%%%%%%%%%%%%%%%%%%%%%%%%%%%%%%%%%%%%%%%%%%%%%%%%%%%%%%%%%%%%%%%%%%
\section{The eddy viscosity}
%%%%%%%%%%%%%%%%%%%%%%%%%%%%%%%%%%%%%%%%%%%%%%%%%%%%%%%%%%%%%%%%%%%%%%%%%%%%%%%%%%%%%%%%%%%%%%%%%%%%%

\begin{figure*}[htb]
\begin{center}
\includegraphics[width=7.0cm,height=5.0cm]{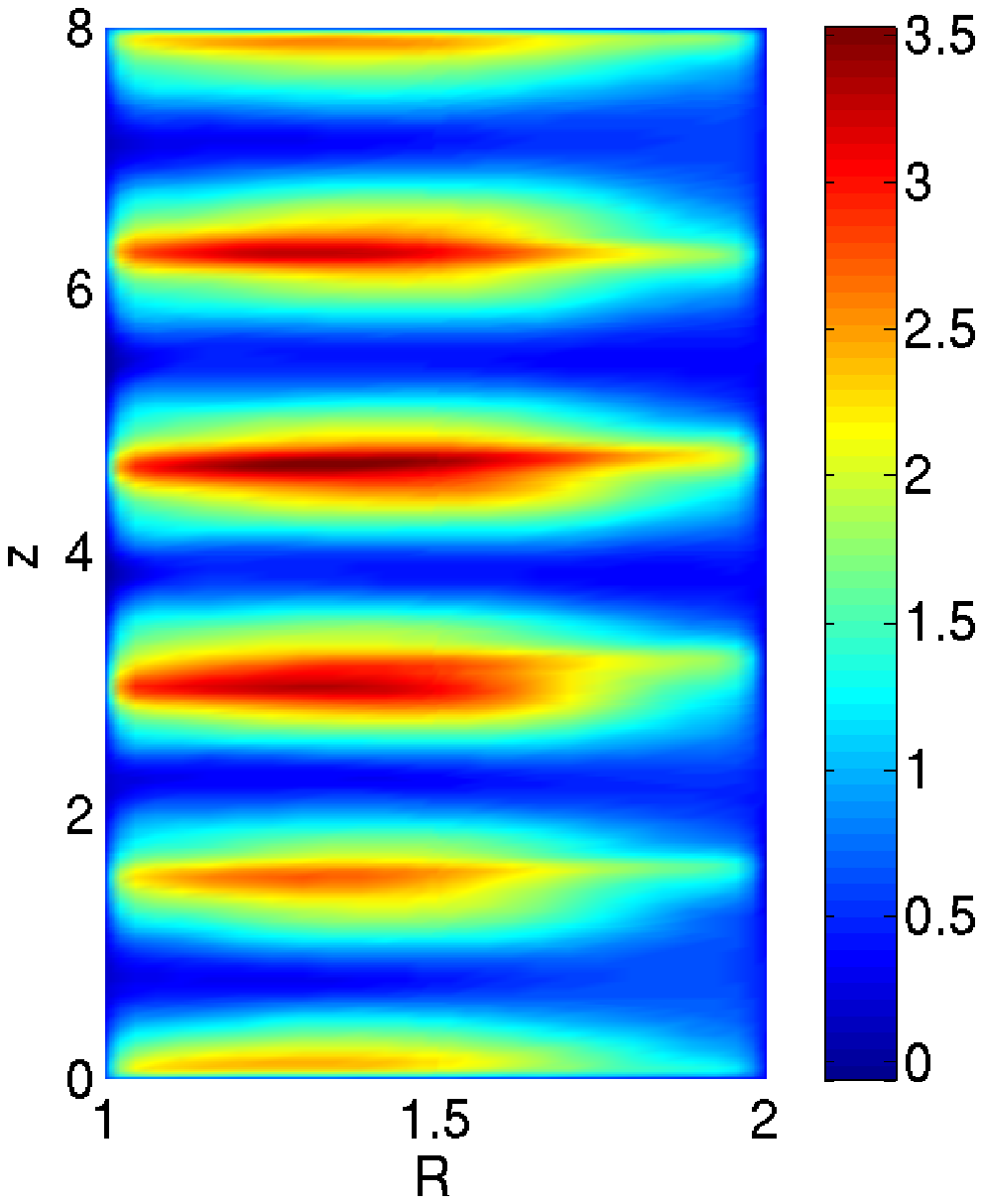}
\hspace*{1cm}
\includegraphics[width=7.0cm,height=5.0cm]{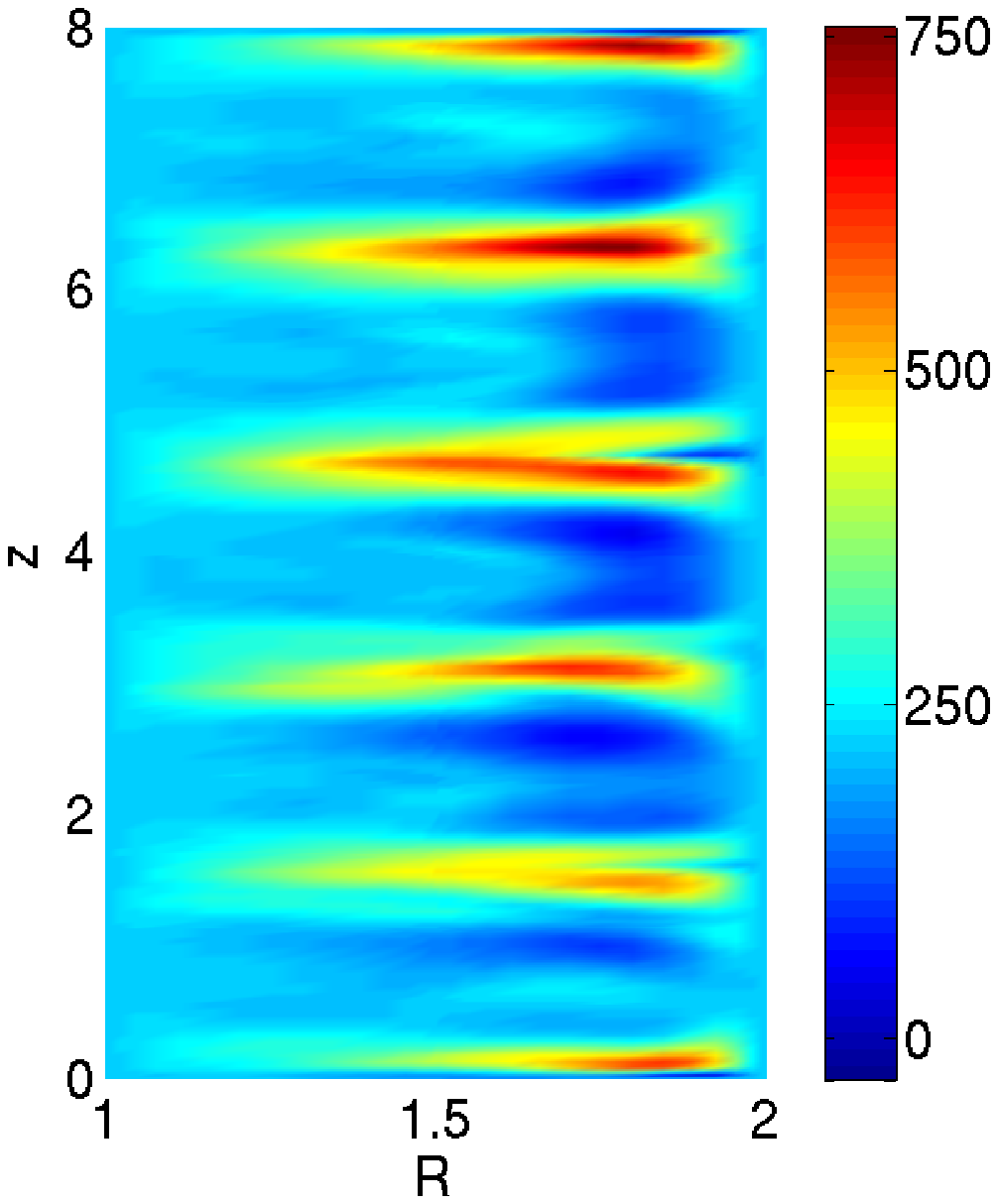}
\caption{\label{fig7} $\Ha=200$, $\Re=500$. Simulations for a flow with $\Pm=0.01$, $\muo=0.35$ and $\mu_B=1$.
        Left: $\eta_{\rm T}/\eta$, right: $\nu_{\rm T}/\nu$. The resulting magnetic
        Prandtl number $\nu_{\rm T}/\eta_{\rm T}$ is about 2.1.}
\end{center}
\end{figure*}

Experiments with  Tayler instability under the presence of differential rotation can also provide  eddy viscosity 
measurements due to the angular momentum transport by both Reynolds stress and Maxwell stress. Within the diffusion 
approximation it is 
 \begin{equation}
 T =\langle u_R' u_\phi'\rangle -  \frac{\langle B_R' B_\phi'\rangle}{\mu_0\rho} = - \nu_{\rm T} R \frac{{\rm d}\Omega}{{\rm d}R}
\label{T2}
\end{equation}
for the torque in the fluid. 
The fluctuations of flow and field can be calculated with the code. The patterns for the instability-induced diffusivity 
values  for $\rm Pm=0.01$ are shown in Fig. \ref{fig7} for the same model as used in Fig \ref{fig5}, i.e.  
$\muo=0.35$, $\Ha =200$ and $\Re=500$. One finds the turbulence-originated  increase of the eddy viscosity 
$\nu_{\rm T}/\nu$ much larger than for the turbulent diffusivity. The turbulent magnetic Prandtl number  
$\rm Pm_{\rm T}=\nu_{\rm T}/\eta_{\rm T}$ becomes  about 2.05. Similar calculations  for $\Pm=0.1$ lead to  $\rm Pm_{\rm T}=0.71$ while for  $\Pm=1$ the smaller value   0.65 results. The results only weakly depend on the averaging procedure. The given numbers follow after 
averaging over the whole cylinder.  
The turbulent magnetic Prandtl number slightly increases with decreasing 
microscopic $\Pm$; and for small Pm it  reaches values larger than unity. Note the differing results of simulations with 
Pm much smaller than unity and those with $\rm Pm\lsim 1$, \cite{YOUS03}. 

Small $\Pm$ are shown to produce large turbulent values
$\rm Pm_{\rm T}$. We cannot provide results  for $\Pm$ smaller than $0.01$ so far. Only  laboratory experiments 
utilizing Tayler instability in liquid metals with their very small magnetic Prandtl number are able to show whether 
the obtained trend is a general one.
 
%\medskip

\end{document}